\documentclass[12pt,reqno]{amsart}
\usepackage{amsfonts}
\usepackage{mathrsfs}
\usepackage{graphicx} 
\usepackage{epsfig}
\usepackage{amsmath, amsthm}
\usepackage{amssymb}

\numberwithin{equation}{section}
\date{}

\theoremstyle{plain}

\theoremstyle{definition}

\numberwithin{equation}{section}

\theoremstyle{plain}
\newtheorem{thm}{Theorem}[section]
\newtheorem{prop}[thm]{Proposition}
\newtheorem{lem}[thm]{Lemma}
\newtheorem{cor}[thm]{Corollary}

\theoremstyle{definition}

\newtheorem{rem}[thm]{Remark}
\newtheorem{defn}[thm]{Definition}
\newtheorem{eg}[thm]{Example}
\newtheorem{subtitle}[thm]{}
\newtheorem{ex}{Exercise}[section]
\numberwithin{equation}{section}

\def\a{\alpha}
\def\b{\beta}

\def\D{\triangle}

\def\g{\gamma}

\def\l{\lambda}

\def\n{\,\vert \,}

\def\w{\omega}

\def\cb{{\mathcal{B}}}
\def\cc{{\mathcal{C}}}

\def\cg{{\mathcal{G}}}

\def\ck{{\mathcal{K}}}
\def\cl{{\mathcal{L}}}
\def\cm{{\mathcal{M}}}
\def\cn{{\mathcal{N}}}

\def\cp{{\mathcal{P}}}

\def\ct{{\mathcal{T}}}

\def\li{\langle}
\def\ri{\rangle}
\def\n{\ \vert\ }

\def\bs{\bigskip}
\def\ms{\medskip}

\def\ni{\noindent}
\def\ti{\tilde}
\def\p{\partial}

\def\diag{{\rm diag}}

\def\C{\mathbb{C}}

\def\R{\mathbb{R} }
\def\Z{\mathbb{Z}}

\newcommand{\beq}{\begin{equation}}
\newcommand{\eeq}{\end{equation}}
\newcommand{\beg}{\begin{eg}}
\newcommand{\eeg}{\end{eg}}
\newcommand{\bthm}{\begin{thm}}
\newcommand{\ethm}{\end{thm}}
\newcommand{\bprop}{\begin{prop}}
\newcommand{\eprop}{\end{prop}}
\newcommand{\bcor}{\begin{cor}}
\newcommand{\ecor}{\end{cor}}
\newcommand{\blem}{\begin{lem}}
\newcommand{\elem}{\end{lem}}
\newcommand{\bca}{\begin{cases}}
\newcommand{\eca}{\end{cases}}
\newcommand{\brem}{\begin{rem}}
\newcommand{\erem}{\end{rem}}
\newcommand{\bpm}{\begin{pmatrix}}
\newcommand{\epm}{\end{pmatrix}}
\newcommand{\bbm}{\begin{bmatrix}}
\newcommand{\ebm}{\end{bmatrix}}
\newcommand{\bvm}{\begin{vmatrix}}
\newcommand{\evm}{\end{vmatrix}}
\newcommand{\bdefn}{\begin{defn}}
\newcommand{\edefn}{\end{defn}}
\newcommand{\bsub}{\begin{subtitle}}
\newcommand{\esub}{\end{subtitle}}
\newcommand{\bex}{\begin{ex}}
\newcommand{\eex}{\end{ex}}
\newcommand{\ben}{\begin{enumerate}}
\newcommand{\een}{\end{enumerate}}

\def\calB{\mathcal{B}}

\def\calN{\mathcal{N}}

\def\R{\mathbb{R}}

\def\R{\mathbb{R}}

\def\C{\mathbb{C}}

\def\Z{\mathbb{Z}}

\def\det{{\rm det \/ }}

\def\calN{\mathcal{N}}
\def\rd{{\rm \/ d\/}}

\def\an1{\hat A_{n-1}^{(1)}}
\def\an2{\hat A_{2n}^{(2)}}
\def\anc2{(\hat A_{2n}^{(2)}, v_n)}

\def\antwo{\hat A_{2n}^{(2)}}
\def\bn1{\hat B_n^{(1)}}
\def\Cn1{\hat C_n^{(1)}}
\def\cnone{\hat C_n^{(1)}}
\def\ccn1{\hat \C_n^{(1)}}

\def\dd1{D_n^{(1)}}
\def\ddn2{D_{n+1}^{(2)}}

\def\k{\kappa}

\def\ans{\hat A_{2n}^\sigma}
\def\ant{\hat A_{2n-1}^{(2)}}

\def\caln{\mathcal{N}}
\def\bh{\setminus}

\def\gone{\hat \cg^{(1)}}

\begin{document}

\title[]
{Geometric aspects of Miura transformations} \today
\author{Changzheng Qu$^*$}\thanks{$^*$Research supported in part by NSF of China under Grant No. 11631007 and Grant. No. 11971251\/}
\address{$^*$School of Mathematics and Statistics\\Ningbo University, Ningbo, Zhejiang, 315211, China. Email: quchangzheng@nbu.edu.cn}
\author{Zhiwei Wu$^\dag$}\thanks{$^\dag$Research supported in part by NSF of China under Grant No. 12271535 and Guangdong Basic and Applied Basic Research Foundation (No. 2021A1515010234)\/}
\address{$^\dag$ School of Mathematics (Zhuhai)\\ Sun Yat-sen University\\ Zhuhai, Guangdong, 519082, China. Email: wuzhiwei3@mail.sysu.edu.cn}


\begin{abstract}
The Miura transformation plays a crucial role in the study of integrable systems. There have been various extensions of the Miura transformation, which have been used to relate different kinds of integrable equations and to classify the bi-Hamiltonian structures. In this paper, we are mainly concerned with the geometric aspects of the Miura transformation. The generalized Miura transformations from the mKdV-type hierarchies to the KdV-type hierarchies are constructed under both algebraic and geometric settings. It is shown that the Miura transformations not only relate integrable curve flows in different geometries but also induce the transition between different moving frames. Other geometric formulations are also investigated.
\end{abstract}

\maketitle \numberwithin{equation}{section}
	

\section{Introduction}

In \cite{MIU68}, Miura introduced the remarkable transformation
\begin{eqnarray}\label{miu}
u=q^2-q_x
\end{eqnarray}
relating solutions of the KdV equation
\begin{eqnarray*}
u_t=u_{xxx}-6uu_x
\end{eqnarray*}
with solutions of the modified KdV (mKdV) equation
\begin{eqnarray*}
q_t=q_{xxx}-6q^2q_x.
\end{eqnarray*}
Nowadays, \eqref{miu} is called the Miura transformation, which has been used to construct  an infinite number of conservation laws of the KdV equation in that time \cite{MGK68}. And the generalized Miura transformation induces B\"{a}cklund transformations for the Gelfand-Dickey hierarchy \cite{A81}.

The Miura transformation adopts various extensions, and they have a number of applications in the study of integrable nonlinear dispersive equations. For examples, \eqref{miu} is equivalent to the spectral problem of the Schr\"{o}dinger operator for the KdV equation, it can be used to study the integrability of the PDE systems. The Miura transformation relates the Hamiltonian structures and conservation laws of the KdV equation with those of the mKdV equation. The Benjamin-Ono equation admits a two-parameter family of Miura transformations, which leads to its infinitely many family of conservation laws \cite{BK79}. The Boussinesq equation is related to dispersive water wave system by the two-component Miura transformations \cite{Kup85}. Moreover, Miura-type transformations connect solutions of  Sawada-Kotera (SK) \cite{SK74} and Kaup-Kupershmidt (KK) equations \cite{Kau80,Kup85} to the solutions of Fordy-Gibbon-Jimbo-Miwa (FGJM) equation \cite{FG80}.

It is worth to point out that, the Miura transformation has been used to study the lower-regularity of the KdV equation by using the result of the mKdV equation \cite{CKST03}. The $L^2$-stability of solitons of the KdV equation was proved from the kink solution of the Gardner equation via the Miura transformation \cite{AMV13, MV03}. Miura transformation admits various extensions, which have been applied to relate different kinds of integrable equations \cite{KLOQ16, LZ11}. Interestingly, the Miura-type transformations can be applied to classify dispersionless integrable systems \cite{DLZ06,DLZ08}. The Miura transformation relating $^L G$-opers to $^L G$-opers on the punctured disc gives an affine analogue of the Harish-Chandra homomorphism obtained by evaluating central elements on the Wakimoto modules \cite{Fre05,FR96}. It was shown in \cite{FL88, FF96,FR96} that the Miura transformations provides homomorphisms of different Poisson algebras. Recently, a direct correspondence between the operators obtained by the Miura transformation and those of the quantum toroidal algebra is found in \cite{HMNW21}.

Several ways have been developed to construct Miura transformations between integrable systems. In a series of papers by Fordy etal. (c.f. \cite{AF89-1,AF89-2,AFL91,For90}), Miura transformations and its multi-component extensions can be constructed by the factorization of energy-dependent operators. Such construction can be utilized to obtain the Miura maps between super integrable systems. In \cite{Yam94}, a direct scheme for constructing Miura transformations is presented, which also works for the discrete integrable systems. In the discrete systems, the bilinear transformation is a powerful method to construct Miura transformations \cite{JRG98}. The symmetry groups was developed to obtain Miura transformations(c.f. \cite{Gut93}, \cite{Gut94}). Miura transformations between any two scalar evolution equations were classified in \cite{CWX06}, which derive new B\"{a}cklund transformations. For certain cases, Miura transformations also can be generated from the gauge transformation \cite{FGLZ21}.

The mKdV equation has been highly involved in the study of differential geometry. In \cite{L77}, Lamb used the mKdV equation to describe motion of  curves with constant torsion. Chern and Tenenblat characterized the mKdV hierarchy as relations between local invariants of certain foliations on a surface of nonzero constant Gauss curvature in \cite{CT81}. Doliwa and Santini \cite{DS94} obtained the mKdV equation from nonstretching evolution of curves in $S^2$. Moreover, the mKdV equation was the equation satisfied by the curvature of chiral shape arc-length preserving closed curves discussed by Goldstein and Petrich in \cite{GP91}.

The main goal of this paper is to explore the geometric aspect of Miura transformations. Our motivation in part comes from the following facts. First of all, it is found that Miura transformations connect with the planar curve flows in some imprimitive geometries in ${\mathbb R}^2$ with the curve flows in the one-dimensional projective space \cite{CQ02}. Secondly, the Miura transformations connect the curvature in certain geometries with that in their sub-geometries. Furthermore, the Miura transformations enclose the rich algebraic structure of integrable systems. In this paper, we will set up a scheme to construct hierarchies of curve flows. Under certain parallel moving frame, the corresponding principle curvatures are solutions to the $\hat \cg^{(1)}$-mKdV and $\hat \cg^{(2)}$-mKdV hierarchies. This will give a geometric explanation of the mKdV-type hierarchies and generalized Miura transformations.

The organization of this paper is as follows. In Section 2, we give a brief discussion on geometric formations of Miura transformation based on the planar curve flows. A review on the construction of the $\hat \cg^{(1)}$-mKdV and $\hat \cg^{(2)}$-mKdV hierarchies is presented in Section 3. And the relation with pseudo-differential operators is discussed. In Section 4, a variety of curve flows are constructed and the relation to mKdV-type hierarchies is discussed. In Section 5, we give an explicit example of the Bousinessq equation to explain our scheme. And  the last section is left for discussion and prospective projects.
\bs

\section{Geometric formulations of Miura transformations}
In this section, we give a brief discussion on the geometric aspect of the Miura transformation.  According to the Erlangen program, for any Lie group $G$ acting locally and effectively on an open set $U$ in the plane so that its group action does not have a common fixed point, there is an associated Klein geometry, which is the theory of geometric invariants of the transformation groups. The Lie algebra $\cg$ of $G$ consisting of all infinitesimal transformation acting on the plane has been classified up to local diffeomorphisms. They are divided into two classes: primitive and imprimitive cases. The corresponding real vector fields have also been classified (cf. \cite{GKO92}, \cite{Olv-2}). The group $G$ is called imprimitive if there exists an invariant foliation in $U$. For the transitive imprimitive Lie algebras, there are eleven types, among which there are six types of imprimitive Lie algebras of vector fields in the plane, including $SL(2)$, $SL'(2)$, $GL(2)$, $SL^2(2)$, $SL(2,k)$ and $GL(2,k) (k\in \Z^+)$. The invariant planar curve flows on those geometries can be projected to curve flows on ${\R P}^1$ \cite{CQ03}.



It indicates that the Miura transformation arises from the relationship between integrable planar curve flows and one-dimensional projective space. Consider planar curve flows in the centro-affine geometry with imprimitive group $GL(2)$. Given $\g(p)\in \R^{2}\bh \{0\}$, such that $\det(\g, \g_p)\neq 0$. Then $(\g, \g_p)$ is a natural moving frame along $\g$. The curve flow $\gamma(t, s)$ is governed by
\begin{equation*}
\gamma_t=U \gamma+W\g_s,
\end{equation*}
where $s$ is the arc-length parameter, defined by
\begin{eqnarray*}
\rd s=\sqrt{\left |\frac {\det(\gamma_p,\gamma_{pp})}{\det(\gamma,\gamma_p)}\right|}\rd p.
\end{eqnarray*}

And the curvature $\kappa$ of $\gamma$ is given by
\begin{eqnarray*}
\kappa=\frac {\det(\gamma,\gamma_{ss})}{\det(\gamma,\gamma_s)}.
\end{eqnarray*}
Then the structure equation for $\g$ is
\begin{align*}
\bpm \g, \g_s\epm_s=\bpm \g, \g_s \epm \bpm 0 & 1 \\ 1 & \k \epm.
\end{align*}
Assume that the flow is arc-length preserving, then $W$ and $U$ satisfy
\begin{eqnarray*}
W_s-\frac 12 \kappa U_s+\frac 12 U_{ss}=0,
\end{eqnarray*}
while the curvature satisfies the following equation,
\begin{eqnarray*}
\kappa_t=\kappa_sW+2U_s+\frac 12 (\kappa_s U_s+\kappa^2U_s-U_{sss}).
\end{eqnarray*}

It is clear to see that the  geometric flow
\begin{equation*}
\gamma_t=-2\kappa\g+(\kappa_s-\frac 12 \kappa^2)\g_s
\end{equation*}
gives the mKdV euation
\begin{eqnarray*}
\kappa_t=\kappa_{sss}-\frac 32 \kappa^2 \kappa_s-4\kappa_s.
\end{eqnarray*}
One can verify by a straightforward computation that this mKdV equation is related to the KdV equation
$$u_t=u_{sss}-6uu_s$$
 by the Miura transformation
\begin{eqnarray*}
u=-\frac{1}{2} \kappa_s+\frac{1}{4} \kappa^2+\frac{2}{3}.
\end{eqnarray*}
It is shown that such Miura transformation relates the mKdV-flow in $GL(2)$ to the KdV-flow in ${\R P}^1$ \cite{CQ03,CIM09}.

Furthermore, it is noticed that the Lie algebra $GL(2)$ is a subalgebra of the Lie algebra $SL^2(2)$ generated by $\{\p_x,x\p_x,x^2\p_x, \p_u, u\p_u, u^2\p_u \}$.
So the geometry $GL(2)$ is a subgeometry of the geometry $SL^2(2)$. Let $\kappa$ and $\phi$ be the curvatures of planar curves respectively in the geometries $SL^2(2)$ and $GL(2)$. Indeed, it is easy to check that their curvatures are related by the Miura transformation \cite{CQ03}
\begin{eqnarray*}
\kappa=2\phi_s+\phi^2.
\end{eqnarray*}

In the end of this section, we show that the Miura transformation connects the integrable curve flows with different moving frames. Consider the  centro-equiaffine curve $\g(x): \R \to \R^2\backslash\{0\}$, with $x$ the centro-equiaffine arc-length parameter, i.e. $\det(\g, \g_x)=1$. And $(\g, \g_x)$ is the centro-equiaffine  moving frame along $\g$ with curvature $u$. It is known that (cf. \cite{CIM09}, \cite{UP95}, \cite{TWa}) if  $\g$ is a solution of the following equation
\beq\label{ae}
\g_t=\frac{1}{4}u_x\g-\frac{1}{2}u\g_x,
\eeq
then $u$ satisfies the KdV equation
\begin{eqnarray*}
u_t=\frac{1}{4}(u_{xxx}-6uu_x).
\end{eqnarray*}

Let $\eta$ be a smooth vector field along $\g$ such that
\begin{align*}
\bca
\det(\g, \eta)=1, \\
\det(\eta_x, \eta)=0.
\eca
\end{align*}
Such $(\g, \eta)$ is called a {\it centro-equiaffine parallel frame} along $\g$.

Set $q=\det(\g, \eta_x)$, by $\det(\eta_x, \eta)=0$, we get
$$u=q^2-q_x,$$
which is the Miura transformation. And the transition matrix between the centro-equiaffine frame and parallel frame is
\begin{equation*}
(\g, \g_x)=(\g, \eta)\bpm 1 & -q \\ 0 & 1 \epm.
\end{equation*}

We call $q$ a {\it principle curvature} of $\g$ w.r.t the parallel frame $(\g, \eta)$. And \eqref{ae} can be written in terms of the parallel frame
\begin{eqnarray*}
\g_t=\frac{1}{4}(2q^3-q_{xx})\g-\frac{1}{2}(q^2-2q_x)\eta,
\end{eqnarray*}
with $q$ satisfying the mKdV equation
\begin{eqnarray*}
q_t=\frac{1}{4}(q_{xxx}-6q^2q_x).
\end{eqnarray*}

Note that if we consider the following third-ordered centro-equiaffine curve flow:
\begin{equation*}
\g_t=\frac{1}{9}(uu_x-u_{xxx})\g+\frac{1}{9}(2u_{xx}-u^2)\g_x,
\end{equation*}
then the equal centro-equiaffine curvature $u$ is a solution to the Sawada-Kotera (SK) equation \cite{SK74}:
\beq\label{sk}
u_t=-\frac{1}{9}(u_{xxxxx}-5uu_{xxx}-5u_xu_{xx}+5u^2u_x).
\eeq
The same Miura transformation \eqref{miu} takes the solution $q$ of Fordy-Gibbons-Jimbo-Miwa (FGJM) equation (cf. \cite{FG80a}, \cite{JM83}):
\begin{equation*}
q_t=-\frac{1}{9}(q_{xxxx}-5q^2q_{xx}-5q_xq_{xx}-5qq_x^2+q^5)_x
\end{equation*}
to the solution of \eqref{sk}.


From the above discussion, we find that the Miura transformation used to obtain an infinite number of conservation laws has natural geometric formulations. Such formulations motivate us to investigate further applications of Miura transformations in other geometric settings.

\bs
\section{The $\hat \cg^{(1)}$-mKdV and $\hat \cg^{(2)}$-mKdV hierarchies}\label{sb}
In this section, we give a brief introduction to the construction of the mKdV-type hierarchies associated to affine Kac-Moody algebras introduced in \cite{DS84} from Lie algebra splittings.

Let  $G$ be a non-compact, real simple Lie group, $\cg$ its Lie algebra, and
 $$\hat \cg^{(1)}=\cl(\cg)=\left\{\sum_{i\leq n_0} \xi_i \l^i \n n_0 \, {\rm an \, integer,\,} \xi_i\in \cg\right\},$$
 the set of smooth loops on $\cg$.

Let
$$\hat\cg^{(1)}_+=\left\{\sum_{i\geq 0}\xi_i\l^i \in \cl(\cg)\right\},\quad \hat\cg^{(1)}_-= \left\{\sum_{i<0} \xi_i\l^i\in \cl(\cg)\right\}.$$
Then $(\hat \cg^{(1)}_+, \hat \cg^{(1)}_-)$ is a splitting of $\hat \cg^{(1)}$.

Let  $\{\a_1, \ldots, \a_n\}$ be a simple root system of $\cg$,  and $\cc$, $\cb_+$, $\cb_-$, $\caln_+$  the Cartan, Borel subalgebras of $\cg$ of non-negative roots, non-positive roots, and positive roots respectively. Let $C$, $B_+$, $B_-$, $N_+$ be  connected subgroups of $G$ with Lie algebras $\cc$, $\cb_+$, $\cb_-$, $\caln_+$ respectively.
Let
\beq\label{ib}
J=\b\l +b,
\eeq
where $b= -\sum_{i=1}^n \a_i$ and $\b$ is the highest root.
\bthm(\cite{DS84}, \cite{TU11}, \cite{TWb}) \label{ta}
Given $q \in C^{\infty}(\R, \cb_+)$, then there exists a unique $S(q,\l)=\sum_{i \leq 1}S_{1, j}(q)\l^j \in \hat\cg^{(1)}$ satisfying
\begin{eqnarray*}
\bca [\p_x+b+q, S(q,\l)]=0,\\
m(S(q,\l))=0, \eca
\end{eqnarray*}
where $m$ is the minimal polynomial of $J$ defined by \eqref{ib}.
\ethm
Assume that there is a sequence of increasing positive integers $\{n_j\n j\geq 1\}$ such that $J^{n_j}$ lies in $\gone_+$ for all $j\geq 1$. Then $\{J^{n_j} \in \gone_+\}$ is called a \emph{vacuum sequence}. Write
$$S^{n_j}(q,\l)= \sum_i S_{n_j, i}(q)\l^i.$$
Then the $n_j$-th flow in the $\gone$-hierarchy for $q:\R^2\to \cb_+$ is
\begin{equation*}
q_{t_{n_j}}= [\p_x+ b+ q, S_{n_j, 0}(q)].
\end{equation*}

The {\it $\hat \cg^{1}$-KdV hierarchy} can be constructed from pushing down the $\hat \cg^{1}$-hierarchy to certain cross-section $C^{\infty}(\R, V)$ along the orbit of the  gauge action of $C^{\infty}(\R, N_+)$ on $C^{\infty}(\R, \cb_+)$  (cf. \cite{DS84}, \cite{TWe}).

Let $\pi_{bn}$ be the projection of $\cg$ onto $\calB_-$ with respect to $\cg=\calB_-\oplus\calN_+$. Then from a direct computation, the $\hat \cg^{(1)}$-hierarchy induces a flow on the space of $C^{\infty}(\R^2, \cc)$, which is called the {\it $\gone$-mKdV hierarchy}.

The $n_j$-th $\gone$-mKdV flow is
\beq\label{il}
q_{t_{n_j}}= [\p_x+ b+ q, \pi_{bn}(S_{n_j, 0}(q))].
\eeq

Let $\sigma$ be a complex linear involution of $\cg$, and $\ck, \cp$ the $1, -1$ eigenspaces of $\sigma$ respectively.

The {\it $\hat \cg^{(2)}$-hierarchy} is constructed from the splitting $(\hat \cg_+^{(2)}, \hat \cg_-^{(2)})$ of $\hat \cg^{(2)}$, where
\begin{align*}
& \hat \cg^{(2)}=\{\xi(\l) \in \cg^{(1)} \mid \overline{\xi(\bar \l)}=\xi(\l), \sigma(\xi(-\l))=\xi(\l)\}, \\
& \hat \cg^{(2)}_+=\hat \cg^{(2)}\cap \hat \cg_+^{(1)}, \quad  \hat \cg^{(2)}_-=\hat \cg^{(2)}\cap \hat \cg_-^{(1)}.
\end{align*}
If there is a simple root system of $\cg$ such that $\b \in \cp$ and $b \in \ck$, then $C^{\infty}(\R, \ck \cap \cc)$ is invariant under \eqref{il}, which induce the {\it $\hat \cg^{(2)}$-mKdV hierarchy}. And the {\it $\hat \cg^{(2)}$-KdV hierarchy} can be constructed in a similar way from the $\hat \cg^{(2)}$-hierarchy as from the $\hat \cg^{1}$-hierarchy to the $\hat \cg^{1}$-KdV hierarchy.

\bdefn (\cite{DS84}) The generalized Miura transformation is the gauge transformation of $C^{\infty}(\R, N_+)$ from the $\hat \cg^{1}$- ($\hat \cg^{(2)}$-, resp.) mKdV hierarchy to the $\hat \cg^{(1)}$- ($\hat \cg^{(2)}$-, resp.) KdV hierarchy.
\edefn

Next we give some explicit examples of the mKdV-type hierarchies from the algorithm given by Theorem \ref{ta} and corresponding Miura-type transformations. Let
\begin{align*}
\calB_n^+&=\{y=(y_{ij})\in sl(n,\C)\n y_{ij}=0, i>j\},\\
\calB_n^-&=\{y=(y_{ij})\in sl(n,\C)\n y_{ij}=0, i < j\},\\
\calN_n^+&=\{y=(y_{ij})\in sl(n,\C)\n y_{ij}=0, i\geq j \},\\
\ct_n &=\{y\in gl(n,\C)\n y_{ij}=0, i \neq j \}
\end{align*}
denote the
subalgebras of upper triangular, lower triangular, strictly upper triangular matrices in $sl(n,\C)$ and diagonal matrices in $gl(n,\C)$ respectively. Let $N_n^+$ denote the corresponding Lie  subgroup of $\calN_n^+$.
\ms
\subsection{The $\hat A_{n-1}^{(1)}$-mKdV hierarchy} \

In the case when $\cg=sl(n)$, $b=\sum_{i=1}^{n-1}e_{i+1, i}$. The Catran subalgebra is $\ct_n$. Let $\pi_{bn}$ be the projection of $sl(n)$ onto $\cb_n^-$ with respect to $sl(n)=\cb_n^-\oplus \cn_n^+$. Let $\pi_{bn}$ be the projection of $sl(n)$ onto $\cb_n^-$ with respect to $sl(n)=\cb_n^-\oplus \cn_n^+$.
The $j$-th $\hat A_{n-1}^{(1)}$-mKdV flow is \eqref{il} for $q=\diag(q_1, \ldots, q_{n}) \in C^{\infty}(\R^2, \ct_n)$,
\beq\label{in}
q_t=[\p_x+b+q, \pi_{bn}(S_{j, 0}(q))].
\eeq

Note that the $\hat A_{n-1}^{(1)}$-KdV hierarchy is the Gelfand-Dickey hierarchy, and the $\hat A_{n-1}^{(1)}$-mKdV hierarchy is the Drinfeld-Sokolov $n \times n$ mKdV hierarchy. In particular, the third $\hat A_{1}^{(1)}$-mKdV flow is the mKdV equation
\begin{equation*}
q_t=\frac{1}{4}(q_{xxx}-6q^2q_x).
\end{equation*}

\ms
\subsection{The $\antwo$-mKdV hierarchy}\

Let $\R^{n+1, n}$ be the linear space $\R^{2n+1}$ equipped with the non-degenerate bilinear form
\beq\label{rho}
\li  X, Y\ri=X^t\rho_n Y,   \quad \rho=\sum_{i=1}^{2n+1}(-1)^{n+i-1}e_{i, 2n+2-i}.
\eeq

Let $O_{\C}(n+1, n)$ be the group of linear isomorphisms of $\C^{2n+1}$ that preserve
$\langle \ , \ \rangle$, i.e.
\begin{equation*}
O_{\C}(n+1, n)=\{g \in SL(2n+1, \C) \mid  g^t\rho_ng=\rho_n\}.
\end{equation*}
Its Lie algebra is  $o_{\C}(n+1, n)=\{A \in sl(2n+1, \C) \mid  A^t\rho_n+\rho_nA=0\}$.
Note that $A=(A_{ij})\in o_\C(n+1, n)$ if and only if
\ben
\item[(i)]
$A_{ij}$'s are symmetric (skew-symmetric resp.) with respect to the skew diagonal line $i+j= 2n+2$ if $i+j$ is odd (even resp.),
\item[(ii)] $A_{ij}=0$ if $i+j= 2n+2$.
\een

Let
\begin{equation}\label{j}
J=e_{1, 2n}\l+\sum_{i=1}^{2n}e_{i+1,i}.
\end{equation}

The $(2j-1)$-th $\antwo$-KdV flow (\cite{TWd}) is an evolution equation for
\beq\label{b}
u=\sum_{i=1}^{n}u_i\b_i, \quad \b_i=e_{n+1-i, n+i}+e_{n+2-i, n+1+i}.
\eeq

In this case, $\pi_{bn}$ is the projection of $o(n+1, n)$ onto $\cb_{2n+1}^-\cap o(n+1, n)$ with respect to
$$o(n+1, n)=(\cb_{2n+1}^-\cap o(n+1, n))\cap (\cn_{2n+1}^+\cap o(n+1, n)).$$
The $(2j-1)$-th $\antwo$-mKdV flow is \eqref{il} for
$$q=\sum_{i=1}^{n}q_i(e_{n+1-i, n+1-i}-e_{n+1+i, n+1+i}) \in C^{\infty}(\R^2, \ct_{2n+1}\cap o(n+1, n)).$$

From a direct computation, we see that the fifth $\hat A_2^{(2)}$-mKdV flow for $\diag(q, 0, -q)$ is the Fordy-Gibbons-Jimbo-Miura (FGJM) equation:
\beq\label{da}
q_t=-\frac{1}{9}(q_{xxxx}-5q^2q_{xx}-5q_xq_{xx}-5qq_x^2+q^5)_x.
\eeq

And the fifth $\hat A_2^{(2)}$-KdV flow is the Kaup-Kupershmidt (KK) equation:
\beq\label{kk}
u_t=-\frac 19 (u_{xxxxx}-10uu_{xxx}-25u_xu_{xx}+2u^2u_x).
\eeq
And the Miura transformation $u=q_x+\frac 12 q^2$ take the solution of \eqref{da} to a solution of \eqref{kk}.

\beg[The Miura transformation from $\hat A_{4}^{(2)}$-KdV to $\hat A_{4}^{(2)}$-mKdV]\

In this case, the Miura transformation is induced by
$$\D_n \in C^{\infty}(\R, N_{2n+1}^+\cap O(n+1, n))$$ such that
\begin{equation*}
\D_n\left(\p_x+\bpm 0 & 0 & 0 & u_2 & \l \\ 1 & 0 & u_1 & 0 & u_2 \\ 0 & 1 & 0 & u_1 & 0 \\ 0 & 0 & 1 & 0 & 0 \\ 0 & 0 & 0 & 1 & 0  \epm\right)\D_n^{-1}=\p_x+\bpm q_2 & 0 & 0 & 0 & \l  \\ 1 & q_1 & 0 & 0 & 0 \\ 0 & 1 & 0 & 0 & 0 \\ 0 & 0 & 1 & -q_1 & 0 \\ 0 & 0 & 0 & 1 & -q_2\epm.
\end{equation*}
The Miura transformation written in terms of $(q_1, q_2)$ is
\begin{eqnarray*}
\begin{aligned}
\bca
u_1 = & q_{1, x}+2q_{2, x}+\frac{1}{2}(q_1^2+q_2^2),\\
u_2  = & q_{2, xxx}+q_2(q_{2, xx}-q_{1, xx})-q_{1, x}(2q_{2, x}+2q_1q_2+q_2^2) \\
& -\frac{1}{2}q_{2, x}(q_{2, x}+2q_1^2)-\frac{1}{2}q_1^2q_2^2.
\eca
\end{aligned}
\end{eqnarray*}
\eeg

\ms
\subsection{The $\ans$-mKdV hierarchy}\

Set
 \beq\label{aa}
S_n=\sum_{i=1}^{2n}(-1)^{i+1}e_{i,2n+1-i}.
\eeq
Let $\sigma$ be the order four automorphism of $sl(2n+1,\C)$ defined by
$$
\sigma(X)=-D_n X^t D_n^{-1}, \quad {\rm where\,\,} D_n=\diag(i, S_n),
$$
and $S_n$ is given by \eqref{aa}. Let $\cg_j$ be the eigenspace of $\sigma$ with respect to the eigenvalue $i^j$ for $0 \leq j \leq 3$. It follows from a direct computation that we have
\begin{align*}
& \cg_0=\bpm 0 & 0\\ 0& sp(2n, \C)\epm, \quad \cg_1=\left\{\bpm 0 & \xi \\ S_n\xi^t & \eta \epm \,\big|\, S_n\eta^tS_n^{-1}=-i\eta\right\},  \\
& \cg_2=\left\{\bpm \C & 0 \\ 0 & \eta \epm\, \big|\, S_n\eta^tS_n^{-1}=\eta\right\}, \quad \cg_3=\left\{\bpm 0 & \xi \\ -S_n\xi^t & \eta \epm\,\big|\,  S_n\eta^tS_n^{-1}=i\eta\right\}.
\end{align*}
We will use the following notation
\begin{equation*}
\cg_0=\left\{\hat y= \bpm 0& 0\\ 0 & y\epm\, \bigg|\, y\in sp(2n, \C)\right\}.
\end{equation*}

Let
\begin{align*}
& \ans=\left\{A(\l)=\sum_{i \leq n_0}A_i \l^i \n A_i \in sl(2n+1,\C),\, \sigma (A(-i\l)) =A(\l)  \right\}, \\
& (\ans)_+=\left\{\sum_{i \geq 0}A_i\l^i \in \ans \right\}, \qquad (\ans)_-=\left\{\sum_{i < 0}A_i\l^i \in \ans \right\}.
\end{align*}
Then $A(\l) \in \ans$ if and only if $A_i \in \cg_j$, where $i \equiv j ({\rm mod\, }  4)$, and $\ans= (\ans)_+\oplus (\ans)_-$ as a direct sum of linear subspaces.

Let
\begin{equation*}
J_A=(e_{1, 2n+1}+e_{2, 1})\l+\sum_{i=2}^{2n}e_{i+1, i} = (e_{1, 2n+1}+e_{2, 1})\l +\bpm 0&0\\ 0& b\epm.
\end{equation*}
Then $J_A^{2j-1}\in \ans$ for all $j\geq 1$.
Let $\pi_{bn}$ be the projection of $\cg_0$ on to $\cg_0 \cap \cb_{2n+1}^-$ with resect to
$$\cg_0=(\cg_0 \cap \cb_{2n+1}^-)\oplus(\cg_0 \cap \cn_{2n+1}^+).$$

It can be checked that $W_n=\oplus_{i=1}^n\R e_{n+2-i, n+1-i}$ is a cross-section of the gauge orbit, and the $\ans$-KdV hierarchy is generated by
\begin{equation*}
\p_x+J_A+\sum_{i=1}^nu_i e_{n+2-i, n+1-i}.
\end{equation*}

The $(2j-1)$-th $\ans$-mKdV flow is \eqref{il} for
$$q=\sum_{i=1}^nq_i(e_{n+2-i, n+2-i}-e_{n+1+i, n+i+i}) \in C^{\infty}(\cg_0 \cap \ct_{2n+1}).$$

\beg\

The fifth $\hat A_{2}^\sigma$-KdV flow for $ue_{23}$ is the Sawada-Kotera (SK) equation \eqref{sk}.

The fifth $\hat A_{2}^\sigma$-mKdV flow for $\diag(0, -q, q)$ is again the FGJM \eqref{da}. Moreover, the Miura transformation connecting the solution $q$ of the FGJM and a solution $u$ of the SK \eqref{sk} is $u=q^2-q_x$.

\eeg
\beg\

The third $\hat A_4^\sigma$-KdV flow is a coupled fifth-order system for $u=u_1e_{3, 4}+u_2e_{2, 5}$:
$$
\bca
u_{1, t}=-2u_{1, xxx}+\frac{6}{5}u_1u_{1, x}+3u_{2, x}, \\
u_{2, t}=-\frac{3}{5}u_{1, xxxxx}+u_{2, xxx}+\frac{3}{5}(u_1u_{1, xxx}+u_{1, x}u_{1, xx}+u_{1, x}u_2-u_1u_{2, x}).
\eca
$$

The Miura transformation from the $\hat A_{4}^\sigma$-mKdV hierarchy to the $\hat A_{4}^\sigma$-KdV hierarchy is
\begin{equation*}
\bca
u_{_1}=  q_1^2+q_2^2-q_{1, x}-3q_{2, x}, \\
u_2=  -2q_{2, xxx}+q_2q_{2, xx}+(q_{1, x}-q_1^2)(q_2^2-q_{2, x}) -q_2(q_{1, xx}-2q_1q_{1, x}).
\eca
\end{equation*}
\eeg

\ms
\subsection{The $\ant$-mKdV hierarchy}\

Let $\R^{2n}$ be the symplectic space with the symplectic form
$$
\w(X, Y)=X^t S_n Y,
$$
where $S_n$ is as defined in \eqref{aa}. $Sp(2n)=\{g \in GL(2n,\R)\n g^tS_n g= S_n\}$  the group of linear isomorphisms of $\R^{2n}$ that preserves $\w$, and
$$sp(2n)=\{A \in sl(2n) \mid A^tS_n+S_nA=0\}$$
is the corresponding Lie algebra of $Sp(2n)$.

Let $\k$ be the involution of $sl(2n, \C)$ defined by $$\k(X)=-S_nX^tS_{n}^{-1},$$ where $S_n$ is as in \eqref{aa}.

Let
\begin{equation*}
\ant=\left\{ A(\l)=\sum_{i \leq m_0}A_i\l^i \mid A_i \in sl(2n, \R), \k (A(-\l))=A(\l) \right\}
\end{equation*}
and
\begin{align*}
&(\ant)_+=\left\{\sum_{i \geq 0}A_i \l^i\in \ant \right\}, \quad
(\ant)_-=\left\{\sum_{i <0}A_i \l^i \in  \ant \right\}.
\end{align*}
Then $((\ant)_+,(\ant)_-)$ is a splitting of $\ant$.

In this case, $J_a=\frac{1}{2}(e_{1, 2n-1}+e_{1, 2n})\l+\sum_{i=1}^{2n-1}e_{i+1, i}$, and $\pi_{bn}$ be the projection of $sp(2n, \C)$ on to $sp(2n, \C)\cap \cb_{2n}^-$ with resect to
$$sp(2n, \C)=(sp(2n, \C)\cap \cb_{2n}^-)\oplus(sp(2n, \C)\cap \cn_{2n}^+).$$

The $(2j-1)$-th $\ant$-mKdV flow is \eqref{il} for
$$q=\sum_{i=1}^nq_i(e_{n+1-i, n+1-i}-e_{n+i, n+i}) \in C^{\infty}(sp(2n, \C) \cap \ct_{2n}).$$
\beg\

The third $\hat A_{3}^{2}$-mKdV flow for $q=q_2(e_{11}-e_{44})+q_1(e_{22}-e_{33})$ is
\begin{eqnarray*}
\bca
q_{2, t}=-4q_{2, xxx}+4(q_{1, x}q_2+q_1^2q_2)_x, \\
q_{1, t}=-\frac{2}{3}(6q_{2, x}q_2+2q_1q_{2, x}-4q_1^2q_2-8q_1q_2^2)_x.
\eca
\end{eqnarray*}

The  third $\hat A_{3}^{2}$-KdV flow for $u=u_1e_{23}+u_2e_{14}$ is
\begin{equation*}
\bca
u_{1, t}= 3u_{2, x}, \\
u_{2, t}=u_{2, xxx}-(u_1u_2)_x.
\eca
\end{equation*}

And the explicit formula of the Miura transformation is
\begin{equation*}
\bca
u_1= 3q_{2, x}+q_{1. x}+q_1^2+q_2^2, \\
u_2= q_{2, xxx}+q_2q_{2, xx}-q_{1, xx}q_2-q_{1, x}(q_{2, x}+q_2^2+2q_1q_2)-q_1^2(q_{2, x}+q_2^2).
\eca
\end{equation*}
\eeg
\ms
\subsection{The $\bn1$-mKdV hierarchy}\

Let  $\bn1$ be the Lie algebra of formal power series $\xi(\l)=\sum_{i\geq n_0} \xi_i\l^i$ with some integer $n_0$ that satisfy
\begin{equation*}
\rho_n \xi(\l)+ \xi(\l)^t \rho_n=0, \quad \overline{\xi(\bar{\l})}=\xi(\l),
\end{equation*}
where $\rho$ is as defined in \eqref{rho}.

 Let $(\bn1)_+$ and $(\bn1)_-$ be the sub-algebras of $\bn1$ defined by
\begin{align*}
& (\bn 1)_+ = \{\xi(\l)=\sum_{i \geq 0} \xi_i \l^i \in \bn 1\},\\
& (\bn 1)_-=\{\xi(\l)=\sum_{i < 0}\xi_i\l^i \in \bn 1\}.
\end{align*}

Let
\begin{equation*}
J_B(\l) =\b \l + b,
\end{equation*}
where
\begin{equation*}
\b=\frac{1}{2}(e_{1, 2n}+e_{2, 2n+1}), \quad b= \sum_{i=1}^{2n}e_{i+1, i}.
\end{equation*}
Note that $J_B^{2j}\not\in \bn1$, $J_{B}^{2j-1} (j \geq 1) \in (\bn 1)_+$, and
\begin{equation*}
 J_B^{2n+1}(\l)=\l J_B(\l).
\end{equation*}

Let $\pi_{bn}$ is the projection of $o(n+1, n)$ onto $\cb_{2n+1}^-\cap o(n+1, n)$ with respect to
$$o(n+1, n)=(\cb_{2n+1}^-\cap o(n+1, n))\cap (\cn_{2n+1}^+\cap o(n+1, n)).$$
The $(2j-1)$-th $\bn1$-mKdV flow is \eqref{il} for
$$q=\sum_{i=1}^{n}q_i(e_{n+1-i, n+1-i}-e_{n+1+i, n+1+i}) \in C^{\infty}(\R^2, \ct_{2n+1}\cap o(n+1, n)).$$
\beg\

The $\hat B_1^{(1)}$-mKdV hierarchy is the same as the mKdV hierarchy, and the third $\hat B_1^{(1)}$-mKdV flow for $\diag(-q, 0, q)$ is
$$q_t=q_{xxx}-q^2q_{x}.$$
The Miura transformation from $\diag(-q, 0, q)$  to $u(e_{12}+e_{23})$ is
\begin{equation*}
u=-q_x+\frac 12 q^2.
\end{equation*}

The Miura transformation in the case of $\hat B_2^{(1)}$-mKdV to the $\hat B_2^{(1)}$-KdV is
\begin{align*}
\bca
u_1=&   2q_{2, x}+q_{1, x}+\frac{1}{2}(q_1^2+q_2^2), \\
u_2= & q_{2, xxx}-q_{1, xx}q_2+q_2q_{2, xx}-2q_{1, x}q_{2, x}-\frac{1}{2}q_{2, x}^2 \\&  -q_1^2q_{2, x} 
-q_{1, x}q_2^2-q_{1, x}q_1q_2-\frac{1}{2}q_1^2q_2^2,
\eca
\end{align*}
where the phase space of the $\hat B_2^{(1)}$-KdV is of the form $u=u_1(e_{23}+e_{34})+u_2(e_{14}+e_{25})$.
\eeg
\ms
\subsection{The $\cnone$-mKdV hierarchy}\


Let
\begin{align*}
&\cnone:= \left\{A=\sum_{i}A_i \l^i \mid A_i \in
sp(2n)\right\}, \\
&(\cnone)_+=\left\{\sum_{i \geq 0}A_i \l^i\in \cnone \right\}, \quad
(\cnone)_-=\left\{\sum_{i <0}A_i \l^i \in  \cnone \right\}, \\
& J_C=e_{1, 2n}\l+\sum_{i=1}^{2n-1}e_{i+1, i}.
\end{align*}

Let $\pi_{bn}$ be the projection of $sp(2n, \C)$ on to $sp(2n, \C) \cap \cb_{2n}^-$ with respect to
$$sp(2n, \C)=(sp(2n, \C) \cap \cb_{2n}^-)\cap(sp(2n, \C) \cap \cn_{2n}^+).$$

The $(2j-1)$-th $\cnone$-mKdV hierarchy is \eqref{il} for
$$q=\sum_{i=1}^nq_i(e_{n+1-i, n+1-i}-e_{n+i, n+i}) \in sp(2n, \C) \cap \ct_{2n}.$$

Note that  the $\hat C_1^{(1)}$-mKdV hierarchy is the $2\times 2$-mKdV hierarchy.

\ms
\subsection{Pseudo-differential operator correspondence}\

Drinfled and Sokolov \cite{DS84} have shown  that there are various KdV-type hierarchies and one mKdV-type associated to each affine Kac-Moody algebra. For example, the $\hat A_{n-1}^{(1)}$-KdV hierarchy with phase space $u=\sum_{i=1}^{n-1}u_ie_{i, n}$ is equivalent to the Gelfand-Dickey hierarchy \cite{Dic03} generated by the pseudo-differential operator
$$L_1=\p^n-u_{n-1}\p^{n-2}-\ldots-u_2\p-u_1.$$
Moreover, the $\hat A_{n-1}^{(1)}$-mKdV hierarchy with phase space $q=\diag(q_1, \ldots, q_n)$ is equivalent to the generalized mKdV hierarchy generated by
$$L_2=(\p-q_n)\cdots (\p-q_1).$$
And the transformation of written $L_1$ into the form of $L_2$ introduces the generalized Miura transformation.

Following a similar argument as in \cite{DS84}, we can get a series of results concerning the pseudo-differential operator correspondence for certain $\hat \cg^{(1)}$- and $\hat \cg^{(2)}$-mKdV hierarchies.
\bprop \
\ben
\item  The $\hat A_{2n}^\sigma$-mKdV hierarchy is equivalent to the KP-type hierarchy generated by
$$L=(\p+q_n)\cdots(\p+q_1)(\p-q_1)\cdots(\p-q_n)\p.$$
\item The $\antwo$-mKdV hierarchy is equivalent to the KP-type hierarchy generated by
$$L=(\p+q_n)\cdots(\p+q_1)\p(\p-q_1)\cdots(\p-q_n).$$
\item The $\cnone$-mKdV hierarchy is equivalent to the reduced KP hierarchy generated by
$$L=(\p+q_n)\cdots(\p+q_1)(\p-q_1)\cdots(\p-q_n).$$
\een
\eprop

\bcor The Miura transformation from the $\cnone$-mKdV hierarchy to the $\cnone$-KdV hierarchy is the same as the one from the $\hat A_{2n}^\sigma$-mKdV hierarchy to the $\hat A_{2n}^\sigma$-KdV hierarchy.
\ecor

\bs
\section{Geometric Miura transformations}\label{sc}
It is known that there are natural curve flows explanation for KdV-type hierarchies in the background of different group actions (cf. \cite{CQ02}, \cite{CQ03}, \cite{MB99}-\cite{MB08c}). In the geometry for curves, the moving frames along curves are induced by the group actions and give the corresponding local differential invariants. If we are able to establish a connection between the group actions and the algebraic structure of the integrable hierarchies, there will be a correspondence between the set of geometric invariants and the phase space of integrable equations.  For example, associated to affine Kac-Moody algebra of type A, B, and C, there are centro-equiaffine, isotropic and Lagrangian curves w.r.t. the group action of $SL(n)$, $O(n+1, n)$ and $Sp(n)$ respectively (cf. \cite{TWb}-\cite{TWe}). In this section, we show that the generalized Miura transformations between the mKdV-type and KdV-type hierarchies discussed in the previous sections can be induced by the transition between moving frames along curve flows.

\ms
\subsection{Centro-equiaffine mKdV curve flow}\

Consider the following submaninfold of centro-equiaffine curves in $\R^n\bh\{0\}$,
\beq
\cm_n(I)=\{\g: I \rightarrow \R^n\bh\{0\}\n \det(\g,\g_x, \ldots, \g_x^{(n-1)})=1\},\;\; I=S^1 \ \text{or}\ \R.
\eeq
Let $\g \in \cm_n(I)$, $g=(\g, \g_x, \ldots, \g_x^{(n-1)})$ and $u=\sum_{i=1}^{n-1}u_ie_{in}$ the central-equiaffine moving frame and centro-equiaffine curvature along $\g$. There is a natural connection between the centro-equiaffine curve flows and the $\hat A_{n-1}^{(1)}$-KdV hierarchy \cite{TWb}.

\bdefn A frame $g \in SL(n, \R)$ is called a parallel frame along $\g$ if there exists smooth functions $q_1, \ldots, q_{n-1}$ along $\g$ such that $ge_1=\g$ and
$$g_x=g(b+\diag(q_1, \ldots, q_{n-1}, -\sum_{i=1}^{n-1}q_i)).$$
\edefn
We call these $\{q_1, \ldots, q_{n-1}\}$ a set of \emph{central-equiaffine principle curvatures} along $\g$.

Write $g=(\g, \eta_2, \ldots, \eta_{n+1})$. If $g=(\g, \ti \eta_2, \ldots, \ti \eta_{n+1})$ is another parallel frame along $\g$ with parallel curvature $\ti q_1, \ldots, \ti q_{n-1}$. Then there exists $C \in C^{\infty}(\R, N_n^+)$ such that
$\ti g=gC$, and
$$\ti q=C^{-1}qC+C^{-1}C_x$$
gives one type of the  B\"{a}cklund transformations for the $\hat A_{n-1}^{(1)}$-mKdV hierarchy.
\bprop
Let $\g \in \cm_n(R)$, and $g$ be a parallel frame along $\g$. If its central-equiaffine principle curvatures $q$ satisfies the $j$-th $\hat A_{n-1}^{(1)}$-mKdV flow \eqref{in}, then
\begin{equation*}
\xi(\g)=gS_{j, 0}(q)e_1 \in T_\g\cm_n(\R).
\end{equation*}
\eprop
\begin{proof} Let $V_n=\sum_{i=1}^{n-1}u_ie_{i, n}$. Since $q$ is a solution of the $j$-th $\hat A_{n-1}^{(1)}$-mKdV flow, there exists $M \in C^{\infty}(\R^2, N_n^+)$, such that
$$M(\p_x+J+q)M^{-1}=\p_x+J+u, u\in C^{\infty}(\R^2, V_n).$$
Then $u$ is the solution $j$-th $\hat A_{n-1}^{(1)}$-KdV flow.  Moreover, let $\ti g$ be the central-equiaffine moving frame of $\g$. Then $\ti g=gM^{-1}$, and $u$ is the central-equiaffine curvature of $\g$. A direct computation implies $S^j(u)=M^{-1}S^j(q)M$. Therefore, the $j$-th $\hat A_{n-1}^{(1)}$-KdV flow can be written as
$$u_{t_j}=[\p_x+b+u, M^{-1}S_{j, 0}(q)M-\zeta_j(u)].$$
In other words, $[\p_x+b+u, M^{-1}S_{j, 0}(q)M-\zeta_j(u)] \in V_n$. Then from the result in \cite{TWb},
$$\xi(\g)=\ti g M^{-1}S_{j, 0}(q)Me_1=gS_{j, 0}(q)Me_1 \in T_\g\cm_n(\R).$$
Since $M\in N_n^+$, $gS_{j, 0}(q)Me_1=gS_{j, 0}(q)e_1$. This proves the Proposition.
\end{proof}
Next we give some examples of the central-equiaffine mKdV flow:
\beg \
\ben
\item  Let $g=(\g, \eta)$ be the parallel frame for $\g \in \cm_2$ and $q$ its parallel curvature. The third flow on $\cm_2$ is
\begin{equation*}
\g_t=(\frac{1}{2}q^3-\frac{1}{4}q_{xx})\g-\frac{1}{2}(q^2-q_x)\eta.
\end{equation*}
And the parallel curvature $q$ satisfies the mKdV equation:
$$q_t=\frac{1}{4}(q_{xxx}-6q^2q_x).$$
\item For general $n \geq 3$, let $g=(\g, \eta_2, \eta_3, \ldots, \eta_{2n+1})$ and
$$q=\diag\left(q_1, \ldots, q_{n-1}, -\sum_{i=1}^{n-1}q_i\right)$$ be the parallel frame and curvature respectively. The second mKdV central-equiaffine curve flow is
\begin{align*}
\g_t & =-\frac{1}{n}((2n-1)q_{1, x}+(n-1)q_1^2+\sum_{i=2}^{n-1}((n-i)q_{i, x}-q_i(\sum_{j=1}^{i}q_j)))\g \\
& \quad +(q_1+q_2)\eta_2+\eta_3.
\end{align*}
These $q_i$'s satisfy the second $\hat A_{n-1}^{(1)}$-mKdV flow.
\een
\eeg
\ms
\subsection{Isotropic mKdV flow} \

Let $\g$ be an isotropic curve in $\R^{n+1, n}$, and $u_1, \ldots, u_{n}$ its isotropic curvatures. That is,
\ben
\item $\{ \g, \g_x, \g_x^{(n-1)} \}$ forms a maximal isotropic subspace of $\R^{n+1, n}$.
\item $\li \g_x^{(n)}, \g_x^{(n)} \ri$=1.
\item There exists unique $\ti g=(\g, \g_x, \ldots, \g_x^{(n)}, p_{n+2}, \ldots, p_{2n+1})$ such that
$$
\ti g_x=\ti g(\sum_{i=1}^{2n}e_{i+1, i}+\sum_{i=1}^n u_i\b_i)=\ti g(b+u),
$$
where $\b_i$ is as define in \eqref{b}.
\een

It is known that there are two types of isotropic curve flows associated to the $\hat A_{2n}^{(2)}$-KdV and $\hat C_n^{(1)}$-KdV hierarchies respectively (cf. \cite{TWd}).

A {\it parallel frame} $g$ along $\g$ is for $g=(\g, \eta_2, \ldots, \eta_{2n+1}) \in O(n+1, n)$ such that
\begin{equation*}
g_x=g(b+\diag(q_n, \ldots, q_{1}, 0, -q_1, \ldots, -q_{n}))
\end{equation*}
for some smooth functions $q_1, \ldots, q_n$ along $\g$. These $q_i$'s are called {\it isotropic principle curvatures} along $\g$. Let $\ti g$ be the isotropic moving frame along $\g$, and $u_1, \ldots, u_{n}$ its isotropic curvatures. That is
\begin{align*}
& \ti g=(\g, \g_x, \dots, \g_{x}^{(n)}, p_{n+2}, \ldots, p_{2n+1}), \\
& \ti g_x=\ti g(\sum_{i=1}^{2n}e_{i+1, i}+\sum_{i=1}^n u_i\b_i)=\ti g(b+u),
\end{align*}
where $\b_i$ is as define in \eqref{b}.

Then there exist $M \in N_{2n+1}^+$ such that $\ti g=gM$, and
\begin{equation*}
M^{-1}(b+q)M+M^{-1}M_x=b+u.
\end{equation*}
This also gives the Miura transformation from the $\hat A_{2n}^{(2)}$-mKdV hierarchy to the $\hat A_{2n}^{(2)}$-KdV hierarchy.
\beg[Isotropic mKdV curve flow of $A$-type] \
\ben
\item Let $g=(\g, \eta, \eta_x)$ be a parallel frame along isotropic curve $\g \in \R^{2, 1}$, with principle curvatures $q$. The fifth isotropic mKdV flow is
\begin{align*}
\g_t & =\frac{1}{9}(-q_{xxxx}+5q_xq_{xx}+5qq_x^2+5q^2q_{xx}-q^5)\g \\
& \quad +\frac{1}{9}(q_{xxx}-3q_x^2+qq_{xx}-4q^2q_x-q^4)\eta.
\end{align*}
The principle curvature $q$ satisfies the FGJM equation \eqref{da}.
\item Let $g=(\g, \eta_2, \eta_3, \eta_4, \eta_5)$ be a parallel frame along isotropic curve $\g \in \R^{3,2}$, and $q_1, q_2$ its principle curvatures. The third isotropic mKdV flow is
{\small
\begin{align*}
\g_t & = -\frac{1}{5}(q_{2, xx}+3q_{1, xx}+12(q_1^2+q_2^2)_x+6q_{1, x}q_2+3q_1^2q_2+3q_1^3-2q_2^3)\g \\
& \quad -\frac{1}{5}(2q_{2, x}+q_{1, x}-2(q_1^2+q_2^2)+5q_1q_2)\eta_2-(q_1+q_2)\eta_3+\eta_{3, x}.
\end{align*}
}
\een
\eeg

\beg[Isotropic mKdV curve flow of $B$-type]\
\ben
\item Let $g=(\g, \eta, \eta_x)$ be a parallel frame along isotropic curve $\g \in \R^{2, 1}$, and $q$ the principle curvature along $\g$. The third isotropic mKdV curve flow of $B$-type is
\begin{equation*}
\g_t=(\frac{1}{3}q^3-q_{xx})\g+(\g_x-\frac{1}{2}q^2)\eta.
\end{equation*}
\item Let $g=(\g, \eta_2, \eta_3, \eta_4, \eta_5)$ be a parallel frame along isotropic curve $\g \in \R^{3,2}$, and $q_1, q_2$ its principle curvatures. The third isotropic mKdV flow of $B$-type is
\begin{align*}
\g_t & =(q_{2, xx}-\frac{3}{2}q_{1, x}q_2-\frac{3}{4}q_1^2q_2+\frac{1}{4}q_2^3)\g \\
& \quad -\frac{1}{4}(4q_{2, x}+2q_{1, x}-3q_1q_2-2q_2^2-q_1^2)\eta_2+(q_1+q_2)\eta_3+\eta_{3, x}.
\end{align*}
\een
\eeg

\ms
\subsection{Lagrangian mKdV curve flow}\

\bdefn\label{gw}(\cite{TWe})
\ben
\item A linear subspace $V$ of $\R^{2n}$ is {\it isotropic\/} if $\w(x,y)=0$ for all $x,y\in V$. A maximal isotropic subspace has dimension $n$, and is called  {\it Lagrangian\/}.
\item A smooth map $\g:\R\to\R^{2n}$ is a {\it Lagrangian curve\/} if
\ben
\item $\g(s), \g_s(s), \ldots, \g_s^{(2n-1)}(s)$ are linearly independent for all $s\in \R$,
\item the span of $\g(s), \ldots, \g_s^{(n-1)}(s)$ is a Lagrangian subspace of $\R^{2n}$
for all $s\in \R$.
\een
\item $\cm_{2n}=\left\{\gamma \in  \R^{2n} \mid  \g \text{ is Lagrangian}, \, \omega(\g_s^{(n)},
\g_s^{(n-1)})=(-1)^{n}\right\}$.
\een
\edefn

A parallel frame $g$ for a Lagrangian curve $\g \in \cm_{2n}$ on symplectic space $(\R^{2n}, \w)$ is $g=(\g, \eta_2, \ldots, \eta_{2n}) \in Sp(2n)$ such that
\begin{equation*}
g_x=g(b+\diag(q_n, \ldots, q_1, -q_1, \ldots, -q_n)),
\end{equation*}
and $q_1, \ldots, q_n$ are called the {\it Lagrangian principle curvatures} along $\g$.

\beg[Lagrangian mKdV curve flow of $A$-type]\

The third Lagrangian mKdV curve flow of $A$-type on $\R^{2, 2}$ is
\begin{eqnarray*}
\begin{aligned}
\g_t= 4(q_1^2q_2+q_{1, x}q_3-q_{2, xx})\g +4(q_{2, x}-q_1q_2)\eta_2-4q_2\eta_3+\eta_4.
\end{aligned}
\end{eqnarray*}
\eeg



\beg[Lagrangian mKdV curve flow of $C$-type]\

The third Lagrangian mKdV curve flow of $C$-type on $\R^{2, 2}$ is
\begin{align*}
\g_t & =-\frac{1}{8}(q_{2, xx}+3q_{1, xx}+6q_{1, x}(q_1+q_2)+2q_2(q_1^2-q_2^2))\g \\
& \quad +\frac{1}{4}(q_{1, x}-q_{2, x}+q_1^2+q_2^2+4q_1q_2)\eta_2+q_2\eta_3+\eta_4.
\end{align*}
\eeg
\bs
\section{The Bousinessq hierarchy}
In this section, we give an explicit example of the Bousinessq (or the $\hat A_{2}^{(1)}$-KdV) hierarchy. The second flow of the $\hat A_{2}^{(1)}$-KdV hierarchy is the following system:
\begin{equation}\label{bsq}
\begin{cases}
u_{1, t} = u_{1, xx}-\frac{2}{3}u_{2, xxx}+\frac{2}{3}u_2u_{2, x},\\
u_{2, t} = -u_{2, xx}+2u_{1, x}.
\end{cases}
\end{equation}

It gives rise to solutions of the {\it Boussinesq equation\/}:
\begin{equation*}
u_{2, tt}=-\frac{1}{3}u_{2, xxxx}+\frac{4}{3}u_{2, x}^2+\frac{4}{3}u_2u_{2, xx}.
\end{equation*}

\ni {\bf I.} Let $\g$ be a centro-equiaffine curve on $\R^{3}\bh \{0\}$, and $\Gamma=(\g, \g_x, \g_{xx})$ the centro-equiaffine moving frame along $\g$ with $u_1, u_2$ the centro-equiaffine curvature. That is
\begin{equation*}
(\g, \g_x, \g_{xx})_x=(\g, \g_x, \g_{xx})\bpm  0 & 0 & u_1 \\ 1 & 0 & u_2 \\ 0 & 1 & 0\epm.
\end{equation*}
Note that if $\g$ satisfies the equation:
\begin{equation*}
\g_t=-\frac{2}{3}u_2\g+\g_{xx}.
\end{equation*}
Then $(u_1, u_2)$ is a solution of \eqref{bsq} (cf. \cite{CIM13}).

Consider the following centro-equiaffine parallel frame $g=(\g, \eta_2, \eta_3)$ with principle curvature $q_1$, $q_2$:
\begin{equation*}
(\g, \eta_2, \eta_3)_x=(\g, \eta_2, \eta_3)\bpm q_1 & 0 & 0 \\ 1 & q_2 & 0 \\ 0 & 1 & -q_1-q_2 \epm
\end{equation*}

Note that
\begin{equation*}
\Gamma=g\bpm 1 & q_1 & q_{1, x}+q_1^2 \\ 0 & 1 & q_1+q_2 \\ 0 & 0 & 1 \epm.
\end{equation*}

This induces the Miura transformation from the principle curvatures $q_1, q_2$ to the central-equiaffine curvature $u_1$ and $u_2$:
\begin{equation*}
\bca
u_1=q_{1, xx}+q_{1, x}q_1-q_1q_{2, x}-(q_1+q_2)q_1q_2, \\
u_2=2q_{1, x}+q_{2, x}+q_1^2+q_1q_2+q_2^2.
\eca
\end{equation*}

\ni {\bf II.} Consider the following bi-linear form $\li \ , \ \ri$ on $\R^3$:
$$\li X, Y \ri=X^t\rho_1 Y, \quad \rho_1=e_{22}-e_{13}-e_{31}.$$
If $\li \g, \g \ri=1$, and $\li \g_x, \g_x \ri=1$. Then there exist unique $p_3 \in \R^{3}$ satisfying
\begin{align*}
(\g, \g_x, p_3) \in O(2, 1).
\end{align*}
This is the isotropic moving frame of $\g$, and the structure equation is
\begin{equation*}
(\g, \g_x, p_3)_x=(\g, \g_x, p_3)\bpm  0 & u & 0 \\ 1 & 0 & u \\ 0 & 1 & 0\epm.
\end{equation*}

If $\g$ is solution of
$$\g_t=-\frac{1}{9}(u_{xxx}-8uu_x)\g+\frac{1}{9}(u_{xx}-4u^2)\g_x,$$
then $u$ satisfies the KK equation \eqref{kk}.

On the other hand, let $(\g, \hat\eta_2, \hat \eta_3)$ be an isotropic parallel frame for $\g$, which means
\begin{equation*}
(\g, \hat\eta_2, \hat \eta_3)_x=(\g, \hat\eta_2, \hat \eta_3)\bpm q & 0 & 0 \\ 1 & 0 & 0 \\ 0 & 1 & -q \epm.
\end{equation*}
Then the isotropic moving frame and the isotropic parallel frame are related by the following gauge transformation:
\begin{equation*}
(\g, \g_x, p_3)=(\g, \hat \eta_2, \hat\eta_3)\bpm 1 & q & \frac{1}{2}q^2 \\ 0 & 1 & q \\ 0 & 0 & 1\epm.
\end{equation*}

So far, we can obtain the Miura transformations among the isotropic curvature ($u$), isotropic principle curvature ($q$), and the centro-equiaffine curvatures ($u_1, u_2$) by the following formula:
\begin{equation*}
u=\frac{1}{2}u_2= q_x+\frac{1}{2}q^2.
\end{equation*}
And the isotropic condition induces the following  reduction on central-equiaffine curvatures and principle curvatures:
$$u_1=\frac{1}{2}u_{2, x}, \quad q_2=0.$$

\bs
\section{Concluding remarks}

In this paper, we present a scheme to study the geometric aspect of the Miura transformations among integrable hierarchies. In general, a transitive group action induces a natural frame along the space of curves that are invariant under the group. From differential geometry, there exists a set of parallel frames along such curves. The transformations from certain parallel frame to the frame induced by the group action naturally generate the generalized Miura transformation from the $\hat \cg^{(1)}$ ($\hat \cg^{(2)}$)-mKdV hierarchy to the $\hat \cg^{(1)}$ ($\hat \cg^{(2)}$)-KdV hierarchy associated. It also turns out that the Miura transformations set up the correspondence between invariant geometric flows in different geometries.

From algebraic structure of the mKdV-type hierarchies present in this paper, it is promising to set up corresponding factorization theory and construct Darboux transformations. Via the Miura transformation, it will give the Darboux transformations for the KdV-type hierarchies. Also being benefited from the algebraic structure, we are able to write down the explicit form of the differential operators which generated the mKdV-type hierarchies as invariant submanifolds of the Kadomtsev-Petviashvili (KP) hierarchy.

In an intriguing paper due to Olver and Rosenau \cite{OR96}, the tri-Hamiltonian duality approach was used to generate several kinds of Camassa-Holm-type equations or systems. It has been shown that the Liouville transformations can set up the correspondences between the hierarchies of the classical integrable equations and those of Camassa-Holm-type equations \cite{KLOQ16, KLOQ17, Mck}. Using the Miura transformations between classical integrable and the Liouville correspondences, one is able to obtain the generalized Miura transformations for Camassa-Holm type equations. However, the geometric formulation of the generalized Miura transformations is not clear, and should be investigated further.


\vskip 0.2cm


\begin{thebibliography}{99}


\bibitem{A81}Adler, M., \emph{On the B\"acklund transformation for the Gel'fand-Dickey equations}, Comm. Math. Phys. \textbf{80(4)} (1981), 517--527.

\bibitem{AMV13}Alejo, M.A., Munoz, C., Vega, L., \emph{The Gardner equation and the $L^2$-stability of the $N$-soliton solution of the Kortweg-De Vries equation}, Trans. Amer. Math. Soc., \textbf{365} (2013), 195-212.

\bibitem{AF89-1} Antonowicz, M., Fordy, A.P., \emph{Factorisation of energy dependent Schr\"{o}dinger operators: Miura maps and modified systems}, Comm. Math. Phys. \textbf{124} (1989), 465-486.

\bibitem{AF89-2}Antonowicz, M., Fordy, A.P., \emph{ Super-extensions of energy dependent Schr\"{o}dinger operators},  Comm. Math. Phys. \textbf{124} (1989), 487-500.

\bibitem{AF90}Antonowicz, M., Fordy, A.P.,  \emph{Hamiltonian structure of nonlinear evolution equations},  Soliton theory: A survey of results, 273-312, Nonlinear Sci. Theory Appl. Manchester Univ. Press, Manchester, (1990).


\bibitem{AFL91}Antonowicz, M., Fordy, A.P.,  Liu, Q.P., \emph{Energy-dependent third-order Lax operators},  Nonlinearity \textbf{4} (1991), 669-684.


\bibitem{BK79}Bock, T., Kruskal, M., \emph{A two-parameter Miura transformation of the Benjamin-Ono equation}, Phys. Lett. A \textbf{74} (1979), 173-176.

\bibitem{CIM09}Calini, A., Ivey, T., Mar\'{i} Beffa, G., \emph{Remarks on KdV-type flows on star-shaped curves}, Physica D \textbf{238} (2009), 788-797.

\bibitem{CIM13}Calini, A., Ivey, T., Mar\'{i} Beffa, G., \emph{Integrable flows for starlike curves in centroaffine space}, Symmetry, Integrability and Geometry: Methods and Applications SIGMA \textbf{9} (2013), 022, 21 pages.

\bibitem{CWX06} Cao, X.F., Wu, H.Y., Xu, C.Y., \emph{On Miura transformations among nonlinear partial
differential equations}, J. Math. Phys. \textbf{47} (2006), 083515.


\bibitem{YC92}
Cheng, Y., \emph{Constraints of the Kadmotesev-Petviashvili hierarchy}, J. Math. Phys., \textbf{33} (1992), 3774-3782.

\bibitem{CT81}
Chern, S.S., Tenenblat, K., \emph{Foliations on a surface of constant curvature and the modified Korteweg-de Vries equations}, J. Differential Geometry \textbf{16} (1981), 347-349.

\bibitem{CQ02}Chou, K.S., Qu, C.Z.,\emph{Integrable equations arising from motions of plane curves}, Physica D \textbf{162} (2002), 9-33.

\bibitem{CQ03}Chou, K.S., Qu, C.Z., \emph{Integrable equations arising from motions of plane curves II}, J. Nonlin. Sci. \textbf{13} (2003), 487-517.

\bibitem{CKST03} Colliander, J., Keel, M., Staffilani, G., Takaoka, H., Tao, T., \emph{Sharp global well-posedness for KdV and modified KdV on ${\mathbb R}$ and ${\mathbb T}$}, J. Amer. Math. Soc. \textbf{16} (2003), 705-749.

\bibitem{Dic03}Dickey, L. A., \emph{Soliton equations and Hamiltonian systems}, second edition, Advanced Series in Mathematical Physics \textbf{26} (2003), World Scientific Publishing Co. Inc., River Edge, NJ.

\bibitem{DS94}
Doliwa, P., Santini, P.M., \emph{An elementary geometric characterization of the integrable motions of a curve}, Phys. Lett. A \textbf{185} (1994), 373-384.

\bibitem{DS84}
Drinfel'd, V.G., Sokolov, V.V., \emph{Lie algebras and equations of Korteweg-de Vries type},  (Russian) Current problems in mathematics, \textbf{24} (1984),  81-180, Itogi Nauki i Tekhniki, Akad. Nauk SSSR, Vsesoyuz. Inst. Nauchn. i Tekhn. Inform., Moscow.

\bibitem{DLZ06} Dubrovin B., Liu, S.Q., Zhang, Y.J., \emph{On Hamiltonian perturbations
of hyperbolic systems of conservation laws I: Quasi-triviality of bi-Hamiltonian perturbations}, Comm. Pure Appl. Math., \textbf{59(4)} (2006),  559-615.

\bibitem{DLZ08} Dubrovin B., Liu, S.Q., Zhang, Y.J., \emph{Frobenius manifolds and central invariants
for the Drinfeld-Sokolov biHamiltonian structures}, Adv. Math., \textbf{219} (2008),  780-837.


\bibitem{FL88}Fateev, V.A., Lukyanov, S., \emph{The models of two-dimensional conformal quantum field theory with $Z_n$ symmetry}, Int. J. Mod. Phys. A \textbf{3} (1988), 507-520.



\bibitem{FF96} Feigin, B., Frenkel, E., \emph{Quantum W-algebras and elliptic algebras}, Comm. Math. Phys. \textbf{178} (1996), 653-678.

\bibitem{FGLZ21}Ferreira, J.C., Gomes, J.F., Lobo, G.V., Zimerman, A.H., \emph{Gauge Miura and B\"{a}cklund transformations for generalized A$_n$-KdV hierarchies},  J. Phy. A: Math. Theor. \textbf{54} (2021), 435201.





\bibitem{For90}Fordy, A.P., \emph{Isospectral flows: their Hamiltonian structures, Miura maps and master symmetries}, in: {\it Solitons in Physics, Mathematics and Nonlinear Optics}, IMA Vol. Math. Appl., \textbf{25}, Springer, New York, (1990), 97-121.

\bibitem{FG80}Fordy, A.P., Gibbons, J., \emph{Some remarkable nonlinear transformations}, Phys. Lett. A \textbf{75} (1980), 325.

\bibitem{FG80a}
Fordy, A.P., Gibbons, J., \emph{Factorization of operators. I. Miura transformations}, J. Math. Phys. \textbf{21} (1980), 2508-2510.

\bibitem{Fre05}
Frenkel, E., \emph{Wakimoto modules, opers and the center at the critical level}, Adv. Math. \textbf{195} (2005),  297-404.

\bibitem{FR96}Frenkel, E., Reshetikhin, N., \emph{Quantum affine algebras and deformations of the Virasoro and W-algebras}, Comm. Math. Phys. \textbf{178} (1996), 237-264.

\bibitem{GH00}Gesztesy, F., Holden, H., \emph{The Cole-Hopf and Miura transformations revisited}, Mathematical physics and stochastic analysis (Lisbon, 1998), 198-214, World Sci. Publ., River Edge, NJ, 2000.

 \bibitem{GRW93}Gesztesy, F., Race, D., Weikard, R.,  \emph{On (modified) Boussinesq-type systems and factorizations of associated linear differential expressions}, J. London Math. Soc., \textbf {47} (1993), 321-340.


\bibitem{GP91}
Goldstein, R.E., Petrich, D.M., \emph{The Korteweg-de Vries hierarchy as dynamics of closed curves in the plane}, Phys. Rev. Lett. \textbf{67} (1991), 3203-3206.

\bibitem{GKO92}
Gonz\'{a}lez-L\'{o}pez, A., Kamran, N., Olver, P., \emph{Lie algebras of vector fields in the real plane}, Proc. London Math. Soc. (3) \textbf{62} (1992), 339-368.

\bibitem{Guh03}Guha, P., \emph{Projective and affine connections on $S^1$ and integrable systems}, J. Geom. Phys. \textbf{46} (2003), 231-242.

\bibitem{Gut93}Guthrie, G. A., \emph{More nonlocal symmetries of the KdV equation}, J. Phys. A Math. Gen. \textbf{26} (1993), L905-L908.

\bibitem{Gut94}Guthrie, G. A., \emph{Recursion operators and non-local symmetries}, Proc. Roy. Soc. London Ser. A \textbf{446} (1994), 107-114.


\bibitem{HMNW21}Harada, K.,  Matsuo, Y., Noshita, G.,  Watanabea, A., \emph{q-deformation of corner vertex operator algebras by Miura transformation}, J. High Energy Phys. \textbf{4} (2021), 202.


\bibitem{JM83} Jimbo, M., Miwa, T., \emph{Solitons and infinite-dimensional Lie algebras}, Publ. Res. Inst. Math. Sci. \textbf{19(3)} (1983), 943-1001.

\bibitem{JRG98} Joshi, N., Ramani, A., Grammaticos, B., \emph{A bilinear approach to discrete Miura transformations}, Phys. Lett. A \textbf{249} (1998), 59-62.

\bibitem{KLOQ16} Kang, J, Liu X.C., Olver P.J., Qu C.Z., \emph{Liouville correspondence between the
modified KdV hierarchy and its dual integrable hierarchy}, J. Nonlinear Sci. \textbf{26} (2016), 141-170.

\bibitem{KLOQ17}Kang, J, Liu X.C., Olver P.J., Qu C.Z.,  \emph{Liouville correspondences between integrable hierarchies}, SIGMA Symmetry Integrability Geom. Methods Appl.  \textbf{13} (2017), 035.

\bibitem{Kau80}Kaup, D.J., \emph{On the inverse scattering problem for cubic eigenvalue problems of the class $\psi_{xxx}+6Q\psi_x+6R\psi=\lambda \psi$},  Stud. Appl. Math. \textbf{62} (1980), 189-216.

\bibitem{Kup84}Kupershmidt, B.A.,  \emph{A super Korteweg-de Vries equation: an integrable system}, Phys. Lett. A \textbf{102} (1984), 213-215.

\bibitem{Kup85}Kupershmidt, B.A., \emph{Mathematics of dispersive water waves}, Comm. Math. Phys. \textbf{99} (1985), 51-73.


\bibitem{KW81}
Kupershmidt, B.A., Wilson, G., \emph{Modifying Lax equations and the second Hamiltonian structure}, Invent. Math. \textbf{62} (1981), 403-436.

\bibitem{L77}
Lamb, G. L., \emph{Solitons on moving space curves}, J. Mathematical Phys. \textbf{18} (1977), 1654-1661.

\bibitem{LZ11}Liu, S.Q., Zhang, Y.J., \emph{Jacobi structures of evolutionary partial differential
equations}, Adv. Math., \textbf{227} (2011), 73-130.


\bibitem{MB99}
Mar{\'{\i}} Beffa, G.,  \emph{The theory of differential invariants and KdV Hamiltonian evolutions}, Bull. Soc. Math. France \textbf{127(3)} (1999), 363-391.

\bibitem{MB08a}
Mar{\'{\i}} Beffa, G., \emph{Projective-type differential invariants and geometric curve evolutions of KdV-type in flat homogeneous manifolds}, Ann. Inst. Fourier, \textbf{58(4)} (2008), 1295-1335.

\bibitem{MB08c}
Mar{\'{\i}} Beffa, G., \emph{Geometric realizations of bi-Hamiltonian completely integrable systems}, SIGMA Symmetry Integrability Geom. Methods Appl. \textbf{4} (2008), Paper 034, 23 pp.

\bibitem{Mck}McKean, H.P., \emph{The Liouville correspondence between the Korteweg-de Vries and the Camassa-Holm hierarchies}, Commun. Pure Appl. Math. {\bf 56} (2003), 998-1015.

\bibitem{MV03}Merle, F., Vega,  L., \emph{$L^2$ stability of solitons for KdV equation},  Int. Math. Res. Not.\textbf{13} (2003), 735-753.

\bibitem{MIU68}
Miura, R., \emph{Korteweg-de Vries equation and generalizations. I. A remarkable explicit nonlinear transformation}, J. Mathematical Phys. \textbf{9} (1968), 1202-1204.


\bibitem{MGK68}
Miura, R., Gardner, C., Kruskal, M., \emph{Korteweg-deVries equation and generalizations. II. Existence of conservation laws and constants of motion}, J. Mathematical Phys. \textbf{9} (1968), 1202-1209.


\bibitem{Olv-1}Olver, P.J., \emph{Applications of Lie Groups to Differential Equations}, second ed., Springer, New York, 1993.
\bibitem{Olv-2}Olver, P.J., \emph{Equivalence, Invariants, and Symmetry}, Cambridge Univ. Press, Cambridge,
1995.

\bibitem{Olv-3}Olver, P.J., \emph{Invariant submanifold flows}, J. Phys. A. Math. Gen. \textbf{41} (2009), 344017.

\bibitem{OR96} Olver, P.J., Rosenau, P., \emph{Tri-Hamiltonian duality between solitons and solitary-wave
 solutions having compact support}, Phys. Rev. E \textbf{53} (1996), 1900-1906.

\bibitem{Pav06}Pavlov, M.V., \emph{The Boussinesq equation and Miura-type transformations}, J. Math. Sci. \textbf{136} (2006), 4478-4483.

\bibitem{UP95}Pinkall, U., \emph{Hamiltonian flows on the space of star-shaped curves},  Results Math. \textbf{27} (1995), 328-332.

\bibitem{SK74} Sawada, K.,  Kotera, t., \emph{ A method for finding N-soliton solutions of the K.d.V. equation and K.d.V.-like equation}, Prog. Theor. Phys., \textbf{51} (1974), 1355-1367.

\bibitem{TU11}
Terng, C.L., Uhlenbeck, K., \emph{The $n\times n$ KdV flows}, J. Fixed Point Theory Appl. \textbf{10} (2011), 37-61.

\bibitem{TU16}
Terng, C.L., Uhlenbeck, K., \emph{Tau function and Virasoro action for the n$\times$n KdV hierarchy}, Comm. Math. Phys. \textbf{342} (2016), 81-116.

\bibitem{TWa}Terng, C.L., Wu, Z., \emph{Central affine curve flow on the plane}, J. Fixed Point Theory Appl., Mme Choquet-Bruhat Fastschrift, \textbf{14} (2013), 375--396.

\bibitem{TWb}Terng, C.L., Wu, Z., \emph{N-dimension central affine curve flows}, J. Differential Geom.\textbf{111} (2019), 145--189.

\bibitem{TWd}Terng, C.L., Wu, Z., \emph{Isotropic curve flows}, Comm. Anal. Geom. \textbf{28} (2020), 1807-1846.

\bibitem{TWe}Terng, C.L., Wu, Z., \emph{Lagrangian curve flows on symplectic spaces}, Symmetry \textbf{13} (2021), 298.

\bibitem{Yam94}Yamilov, R.I., \emph{Construction scheme for discrete Miura transformations}, J. Phys. A: Math. Gen., \textbf{27} (1994), 6839-6851.

\end{thebibliography}
\end{document}